\documentclass[aps,twocolumn,superscriptaddress,showpacs,amsmath,amssymb,longbibliography]{revtex4}
\usepackage{epsfig}
\usepackage{subfigure}
\usepackage{amssymb,amsmath}
\usepackage{graphicx}
\usepackage{epstopdf}
\usepackage{epsfig}
\usepackage{color}
\usepackage{amsfonts}
\usepackage{amscd}
\usepackage{float}
\usepackage[colorlinks,urlcolor=blue,linkcolor=blue,citecolor=blue]{hyperref}
\usepackage{bbm}
\usepackage{mathrsfs}
\usepackage{mathtools}

\emergencystretch=\maxdimen
\DeclareMathSizes{10}{10}{7}{5}

\begin{document}

\title{Frustration Elimination and Excited State Search in Coherent Ising Machines}

\author{Zheng-Yang Zhou}
\affiliation{Zhejiang Key Laboratory of Quantum State Control and Optical Field Manipulation, Department of Physics, Zhejiang Sci-Tech University, 310018 Hangzhou, China}
\affiliation{Theoretical Quantum Physics Laboratory, Cluster for Pioneering Research, RIKEN, Wakoshi, Saitama 351-0198, Japan}
\author{Clemens Gneiting}
\affiliation{Theoretical Quantum Physics Laboratory, Cluster for Pioneering Research, RIKEN, Wakoshi, Saitama 351-0198, Japan}
\affiliation{Center for Quantum Computing, RIKEN, Wakoshi, Saitama 351-0198, Japan}
\author{J. Q. You}
\altaffiliation[jqyou@zju.edu.cn]{}
\affiliation{Zhejiang Key Laboratory of Micro-Nano Quantum Chips and Quantum Control, School of Physics, and State Key Laboratory for Extreme Photonics and Instrumentation, Zhejiang University, Hangzhou 310027, China}
\affiliation{College of Optical Science and Engineering, Zhejiang University, Hangzhou 310027, China}
\author{Franco Nori}
\altaffiliation[fnori@riken.jp]{}
\affiliation{Theoretical Quantum Physics Laboratory, Cluster for Pioneering Research, RIKEN, Wakoshi, Saitama 351-0198, Japan}
\affiliation{Center for Quantum Computing, RIKEN, Wakoshi, Saitama 351-0198, Japan}
\affiliation{Physics Department, The University of Michigan, Ann Arbor, Michigan 48109-1040, USA}

\date{\today}

\begin{abstract}
Frustration, that is, the impossibility of satisfying the energetic preferences between all spin pairs simultaneously, underlies the complexity of many fundamental properties in spin systems, including the computational difficulty in determining their ground states. Coherent Ising machines (CIMs) have been proposed as a promising analog computational approach to efficiently find different degenerate ground states of large and complex Ising models. However, CIMs also face challenges in solving frustrated Ising models: frustration not only reduces the probability of finding good solutions, but it also prohibits the leveraging quantum effects in doing so. To circumvent these detrimental effects of frustration, we show how frustrated Ising models can be mapped to frustration-free CIM configurations by including ancillary modes and modifying the coupling protocol used in current CIM designs. Such frustration elimination may empower current CIMs to benefit from quantum effects in dealing with frustrated Ising models. In addition, these ancillary modes can also enable error detection and searching for excited states.
\end{abstract}

\maketitle
{\it Introduction---}Ising models~\cite{10.1103/RevModPhys.39.883, 10.1038/nature10012, 10.1103/PhysRevLett.121.163601, 10.1038/s42005-021-00792-0, 10.1103/PhysRevResearch.4.013141} have been widely studied, because these models exhibit, despite of their seemingly simple Hamiltonians, a rich variety of interesting properties, exemplified, e.g., by the theory of spin glasses~\cite{10.1103/PhysRevLett.35.1792, 10.1103/PhysRevLett.52.1156, 10.1103/RevModPhys.65.829, 10.1103/PhysRevLett.118.127201, 10.1103/PhysRevX.10.031024}. The price to pay is that the properties of large Ising models with random all-to-all couplings are notoriously difficult to study both experimentally and numerically~\cite{10.1103/PhysRevE.89.062120, 10.1038/nature18274, 10.1088/2058-9565/aa9c59, 10.1126/sciadv.1700791, 10.1103/PhysRevE.106.054143, PhysRevResearch.5.013224}. For instance, finding their ground states is known to be NP-hard.

Coherent Ising machines (CIMs)~\cite{10.1364/OE.19.018091, 10.1103/PhysRevA.88.063853, 10.1038/nphoton.2014.249,10.1126/science.aah5178, 10.1126/science.aah4243,10.1038/nphoton.2016.68, 10.1038/s41534-017-0048-9, 10.1103/PhysRevA.96.053834,arXiv:2204.00276v1,10.1038/s41567-021-01492-w,10.1364/AOP.475823,10.1364/OE.479903,10.1364/OE.426476,10.1038/s41467-023-37065-z} have been developed as optical analog computers with the potential to simulate Ising models more efficiently. These machines feature flexible optical couplings, that can, in principle be adjusted to realize any desired Ising interaction, overcoming the nearest-neighbor restrictions of other hardware approaches. In particular, this allows them to enter deeply into, and explore, the realm of frustrated Ising problems. Frustrated couplings, which are omnipresent in generic Ising models, lie at the heart of many intriguing properties of Ising models, including the computational hardness of finding ground states~\cite{10.1126/science.1064761, 10.1103/PhysRevB.68.060403, 10.1038/nature04447, 10.1038/nature08917, 10.1038/nature09071, 10.1126/science.1207239, 10.1103/RevModPhys.85.1473,10.1103/RevModPhys.86.153}. Therefore, dealing with frustrated Ising models is a core objective of CIMs.

However, frustrated couplings represent challenges as well for CIMs. Most Ising models with frustrated couplings correspond to CIM configurations where loss acts inhomogeneously on different effective optical spins~\cite{10.3390/e18100365, 10.1103/PhysRevLett.122.040607, 10.1126/sciadv.aau0823, 10.1002/qute.202000045, 10.1038/s41467-022-33441-3, 10.1038/s42005-022-00927-x, 10.1103/PhysRevResearch.4.013009}, thus breaking the homogeneous-amplitude requirement of CIMs~\cite{10.1364/OE.19.018091}. Moreover, frustration in CIMs results in intrinsic single-photon loss, which decoheres quantum states~\cite{10.1103/PhysRevLett.120.093601, 10.22331/q-2018-03-27-58, 10.1103/PhysRevResearch.2.043387, 10.1103/PhysRevLett.126.023602, 10.1103/PhysRevA.104.013715,10.1103/PhysRevA.106.023714,10.1103/PhysRevA.108.023704} and thus impedes the advancement of quantum CIMs. Therefore, finding improved ways to deal with frustrated couplings can significantly affect the performance of CIMs.

Although frustration is an intrinsic property of generic Ising models, frustrated optical couplings can be circumvented in CIMs while preserving the to-be-found ground states. Since the energy difference between different spin configurations is mapped to the total-loss difference between optical modes in CIMs~\cite{10.1364/OE.19.018091,10.1038/nphoton.2014.249,10.1038/s41567-024-02420-4}, frustration is absent in CIMs whenever ground states of Ising models are mapped to loss-free optical configurations. Unfortunately, these ``zero points," i.e., the Ising energies corresponding to vanishing optical loss, cannot be adjusted freely in current CIMs.

To solve this obstacle in CIMs, we provide a method to shift the lossless point of CIMs with the help of ancillary modes. In our proposal, the signal degenerate-optical-parametric-oscillator (DOPO) modes express the spin configurations in the same way as in the current CIMs, while the additional ancillas shift the correspondence between Ising energy and optical loss, so that Ising ground states are mapped to lossless optical configurations. Eliminating frustration this way promotes current CIMs mainly in three aspects: First, it allows for the presence of stronger quantum effects in CIMs, e.g., steady states can emerge as superpositions of all the degenerate ground states of the underlying Ising models despite their frustration. Enabling these quantum effects is essential for developing CIMs operating in the quantum regime. Second, inhomogeneities caused by frustration are absent, thus improving the performance of CIMs in the semiclassical regime. Finally, by shifting the lossless point of CIMs to excited energies, it is possible to search for excited states.

{\it Mapping Ising models to CIMs---}In a CIM, the spin states $|\!\uparrow\rangle$ and $|\!\downarrow\rangle$ are expressed by coherent states $|\alpha\rangle$ and $|-\alpha\rangle$, respectively, which are steady states of the uncoupled DOPOs described by the Hamiltonian $H=\sum_kS(a_k^2+{a_k^{\dag}}^2)$ and the two-photon loss $\mathcal{L}_{d}(\rho)=\sum_k\frac{\Gamma_{\rm tp}}{2}(2a_k^2\rho {a_k^{\dag}}^2-{a_k^{\dag}}^2a_k^2\rho-\rho {a_k^{\dag}}^2a_k^2)$. At the same time, Ising coupling terms,
\begin{eqnarray}\label{Ising coupling}
J_{n,m}\sigma_z^{(n)}\sigma_z^{(m)},
\end{eqnarray}
are represented by collective loss channels $\mathcal{L}_{m,n}(\rho)=\frac{\Gamma_c}{2}(2L_{n,m}\rho L^{\dag}_{n,m}-L^{\dag}_{n,m}L_{n,m}\rho-\rho L^{\dag}_{n,m}L_{n,m})$, with the Lindblad operator,
\begin{eqnarray}\label{optical Ising coupling}
L_{n,m}&=&a_n+{\rm sign}(J_{m,n})a_m,
\end{eqnarray}
effectively mapping the energies of different spin configurations to the total loss in the CIM. Here, $\sigma_z^{(n)}$ is the Pauli matrix of the $n$th spin, $J_{n,m} = \pm J$ with $J>0$, $a_n$ is the annihilation operator of the $n$th optical mode in the CIM, and $\rho$ is the density matrix~\cite{SupplementalMaterial}.

For $J_{n,m}>0$, the Ising coupling term (\ref{Ising coupling}) contributes a positive energy if the two spins are aligned, and negative energy if they are antialigned. In the corresponding optical coupling (\ref{optical Ising coupling}), the Lindblad term causes loss if the two optical modes have the same phase $\alpha$, while two optical modes with opposite phases describe a dark state of the loss term. Similarly, if $J_{n,m}<0$, two optical modes with the same phases describe a dark state of the corresponding loss term.

Therefore, the total energy of a spin configuration, i.e., the summed-up contributions of all the Ising coupling terms~(\ref{Ising coupling}), is proportional to the total optical loss in the CIM. In particular, the ground state of the Ising model corresponds to the optical configuration with minimum loss. Note that an optical configuration without loss in a CIM corresponds to a spin configuration with minimum possible energy $E_{\rm MPE}=-JN_{\rm c}$, where $N_{\rm c}$ is the total number of coupling terms~(\ref{Ising coupling}). In such cases, the steady states of CIMs are catlike superposition states in the frustration-free ground-state space~\cite{10.1103/PhysRevA.104.013715,SupplementalMaterial,10.1038/nphys1342}. As the phase transition point (threshold pump power) of a CIM depends on the loss, the optical configuration with minimum loss, which is located at the transition point, can be found by gradually increasing the pump~\cite{SupplementalMaterial}.
\begin{figure}[t]
\center
\includegraphics[width=3.4in]{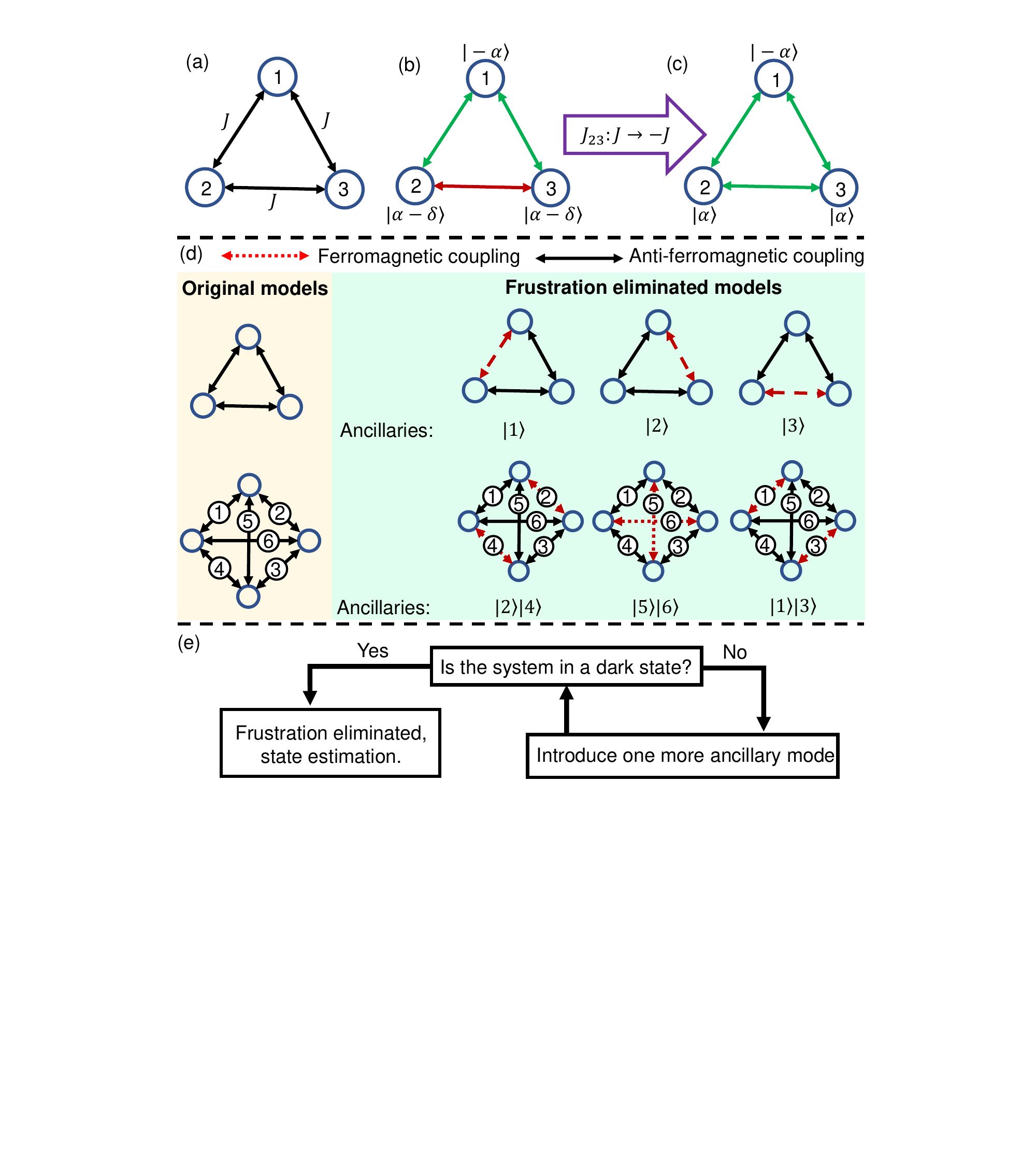}
\caption{Illustration of frustration elimination. (a) Paradigmatic three-mode example featuring frustration, reflected by the absence of a dark state. (b) One of the three degenerate ground states of the configuration in (a), sustaining loss due to the coupling between mode $2$ and mode $3$. The red double-arrow represents the coupling contributing loss, and the green double-arrows represent the loss-free couplings. (c) Flipping the coupling between modes 2 and 3 eliminates frustration for the ground state in (b). (d) Simultaneous frustration elimination for all degenerate ground states with the help of an ancilla. Although frustration in general prohibits dark states in CIMs, it is possible to map all the ground states of a frustrated Ising model to an optical dark state with the help of ancillary modes.}\label{fig1}
\end{figure}

{\it Obstacles caused by frustrated couplings---}For illustration, let us consider the simplest possible situation featuring frustration, realized by three DOPO modes, as shown in Fig.~\ref{fig1}(a). In this example, all three optical couplings energetically prefer coupled modes with opposite phases, which cannot be simultaneously satisfied by any configuration. Therefore, the system unavoidably sustains loss from at least one coupling channel, thus prohibiting quantum superposition and entanglement. The resulting decoherence represents an intrinsic obstacle for developing quantum CIMs.

In addition to suppressing quantum effects, frustration hinders the original function of CIMs, that is, finding ground states of Ising models in the semiclassical regime. In a frustrated CIM configuration, the loss is distributed inhomogeneously among the modes, as illustrated in Fig.~\ref{fig1}(b). Consequently, different modes assume different amplitudes, e.g., mode 1 is above threshold while modes $2$ and $3$ remain below threshold. Such inhomogeneous amplitudes render the mapping in Eqs.~(\ref{Ising coupling}),(\ref{optical Ising coupling}) defective, thus preventing CIMs from finding correct solutions~\cite{SupplementalMaterial}.

In general, Ising models with ground-state energies larger than $E_{\rm MPE}$ are, in the current CIM design, mapped according to Eqs. (\ref{Ising coupling}),(\ref{optical Ising coupling}) to such frustrated couplings.

{\it Frustration elimination: general idea---}For a specific ground state, e.g., $|-\alpha\rangle|\alpha\rangle|\alpha\rangle$, the loss in the frustrated configuration shown in Fig.~\ref{fig1}(b) can be removed by flipping the sign of the lossy coupling term, as shown in Fig.~\ref{fig1}(c). However, such direct manipulation cannot address all the degenerate ground states of the original Ising model, e.g., it fails for the solution $|\alpha\rangle|\alpha\rangle|-\alpha\rangle$.

Therefore, we introduce an ancilla that controls the flipping of the coupling terms. As the ancilla, by design, removes the loss, the steady state can then be a superposition of the different solutions, as illustrated in Fig.~\ref{fig1}(d) and explicitly verified below. In the resulting state, the ancillary part mediates the superposition of three different, unfrustrated Ising models, the ground states of which are encoded in the signal modes. The set of these frustration-free ground states comprises all the degenerate ground states of the original frustrated model.

Figure~\ref{fig1}(d) illustrates how to adapt frustration elimination to general frustrated Ising models~(\ref{Ising coupling}). Flipping an appropriate coupling term, e.g., coupling $5$ in Fig.~\ref{fig1}(d), reduces the energy by at most $2J$ for all spin configurations, and the corresponding ground states of the original model maintain to carry the lowest energy. The lowest energy can be further reduced by flipping several coupling terms, each flip enabled by a separate ancilla as illustrated in Fig.~\ref{fig1}(d), while preserving the ground states. Given a sufficient number of ancillary modes, the energy of the ground states $E_{\rm ground}$ can then be reduced to the minimum possible energy $E_{\rm MPE}$, and the frustration is completely eliminated. Specifically, the number of necessary ancillas is determined by $(E_{\rm ground}-E_{\rm MPE})/(2J)$, which can be identified by either current CIMs or by gradually increasing the number of ancillas.

{\it Ancilla realization in CIMs---}The desired ancillas for frustration elimination can be realized in CIMs by additional optical modes. For a Lindblad operator $(L=a_1+a_2)$ with the dark states $|\alpha\rangle|-\alpha\rangle$ and $|-\alpha\rangle|\alpha\rangle$, the dark states can be modified by including a third mode as
\begin{eqnarray}\label{collective loss modified}
L=a_1+a_2+2a_{\rm an}.
\end{eqnarray}
The collective loss~(\ref{collective loss modified}) supports the dark states
\begin{eqnarray}\label{solutions modified}
|\alpha\rangle|\alpha\rangle|-\alpha\rangle_{\rm an},~~{\rm and}~|-\alpha\rangle|-\alpha\rangle|\alpha\rangle_{\rm an} ,
\end{eqnarray}
where the two signal modes now take the same sign.

To control multiple loss channels with a single ancillary mode, e.g., the case illustrated in Fig.~\ref{fig1}(d), we introduce an effective hyperspin~\cite{10.1038/s41467-022-34847-9,10.1103/PhysRevLett.132.017301}, which can take different directions, as the ancillary mode:
\begin{eqnarray}\label{collective loss with ancillary}
L_{\rm an}=\sum_{n,m}e^{i\phi_{m,n}}\left[a_n+{\rm sign}(J_{m,n})a_m\right]+2a_{\rm an}.
\end{eqnarray}
Different collective coupling terms are now associated with different phases $\phi_{n,m}$, corresponding to rotations of spin directions; while the ancillary mode (hyperspin) autonomously picks up one phase (direction) and flips the corresponding coupling term. An arbitrary phase can be imposed on DOPO modes by a phase and amplitude modulator~\cite{10.1038/nphoton.2014.249}.

Such ancillary states can be realized by nondegenerate optical parametric oscillators (NDOPOs) described by the Hamiltonian
$H_{\rm NDOPO}=S(a_{\rm ani}a_{\rm ans}+a^{\dag}_{\rm ani}a^{\dag}_{\rm ans})$ and the Lindblad operator
$L_{\rm NDOPO}=a_{\rm ani}a_{\rm ans}$ with an ancillary idler mode $a_{\rm ani}$ and an ancillary signal mode $a_{\rm ans}$. The steady states of an NDOPO are also coherent states $|\alpha_{\rm ani}\rangle|\alpha_{\rm ans}\rangle$ with $\alpha_{\rm ani}\alpha_{\rm ans}=2S/\Gamma_{\rm NDOPOtp}$, where $\Gamma_{\rm NDOPOtp}$ is the two-photon loss rate of the NDOPO. Note that an additional loss channel with the same loss rate is required to fix the ``length'' of the hyperspin,
\begin{eqnarray}\label{collective loss with pair ancillary}
L_{\rm ani}=\sum_{n,m}e^{i\phi_{m,n}}[a_n+{\rm sign}(J_{m,n})a_m]+2a_{\rm ani},\nonumber\\
L_{\rm ans}=\sum_{n,m}e^{-i\phi_{m,n}}[a_n+{\rm sign}(J_{m,n})a_m]+2a_{\rm ans},
\end{eqnarray}
with $2S/\Gamma_{\rm NDOPOtp}=|a_n|^2$. To satisfy the two loss channels in Eq.~(\ref{collective loss with pair ancillary}) simultaneously, the two NDOPO modes must have amplitudes with the same absolute value, e.g., $a_{\rm ani}=\exp(i\phi_{m,n})a_n$ and $a_{\rm ans}=\exp(-i\phi_{m,n})a_n$.

\begin{figure}[t]
\center
\includegraphics[width=3.2in]{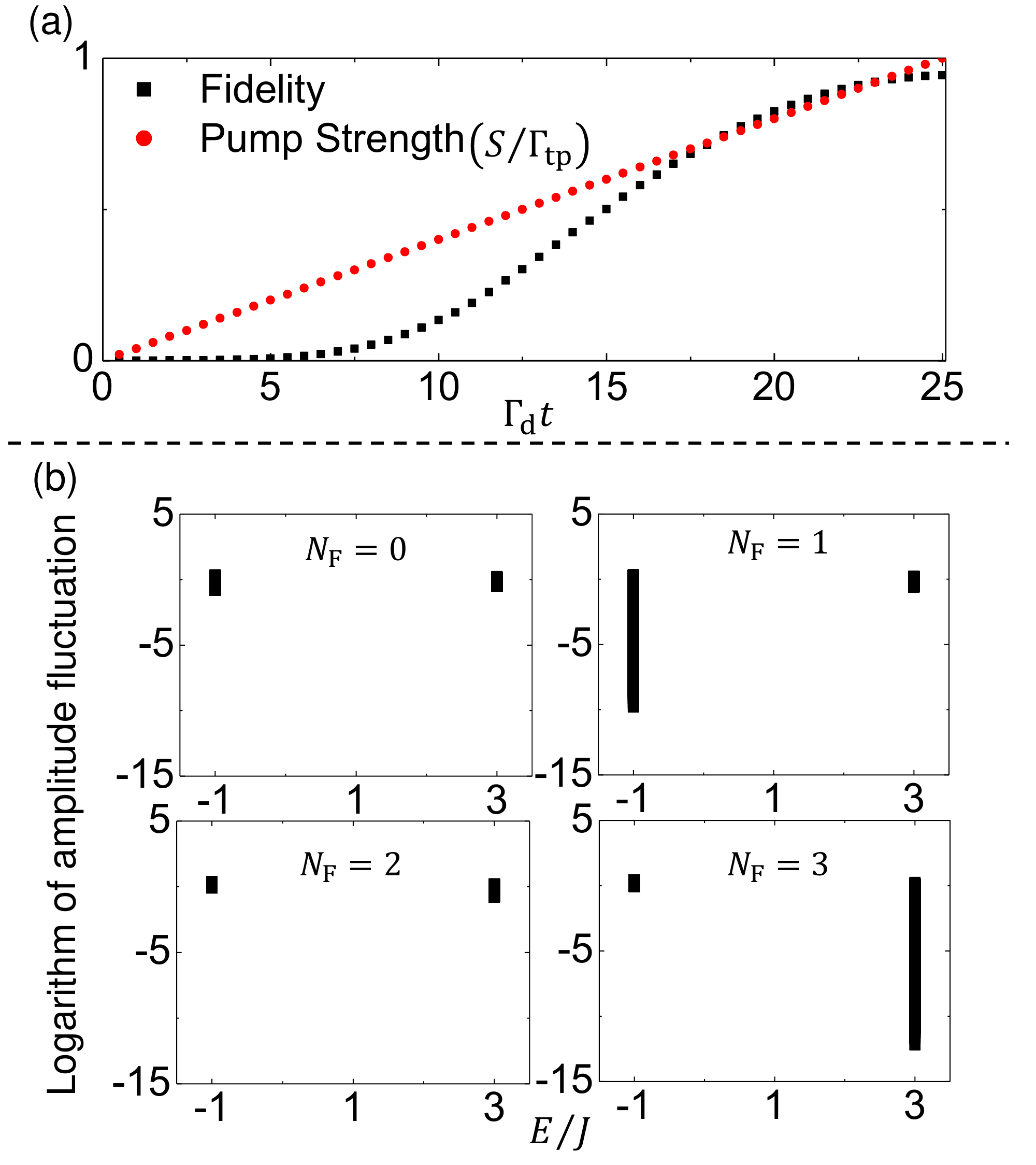}
\caption{Ground-state search under frustration elimination for the frustrated Ising configuration presented in Fig.~\ref{fig1}. Shown is the time-resolved fidelity of the state, generated with a time-dependent pump, with respect to the pure dark state~(\ref{NMdarkstate}). The latter describes a superposition of the six possible ground states. The numerical demonstration of the close reproduction of (\ref{NMdarkstate}) convincingly confirms the validity of the mapping of the frustrated Ising model to an optical effective spin system without frustration. The collective loss rate is $\Gamma_{\rm c}=3\Gamma$, and the two-photon loss rates for the DOPOs and NDOPOs have the same value $(\Gamma_{\rm tp}=\Gamma_{\rm NDOPOtp}=\Gamma)$. (b) Self error detection and arbitrary state searching with frustration elimination method for the model presented in Fig.~\ref{fig1}(a). The correct solutions are indicated by vanishing amplitude inhomogeneity. More active ancillary modes can shift the target state to excited states.}\label{fig2}
\end{figure}
{\it Numerical demonstration of frustration elimination---}We numerically simulate the evolution of the frustration eliminated CIM configuration that corresponds to the frustrated Ising configuration in Fig.~\ref{fig1}(a). The DOPOs and NDOPOs share the same two-photon loss rate $\Gamma_{\rm tp}=\Gamma_{\rm NDOPOtp}=\Gamma$, and the frustration-eliminating loss channels (\ref{collective loss with pair ancillary}) denote three coupling terms with the phases, $\phi_{1,2}=0$, $\phi_{2,3}=\frac{\pi}{2}$, and $\phi_{3,1}=\frac{\pi}{4}$. The three frustration-eliminated configurations in Fig.~\ref{fig1}(d) and the $Z_2$ symmetry result in a six-component dark state,
\begin{eqnarray}\label{NMdarkstate}
|\psi(\infty)\rangle=\frac{1}{\sqrt{6}+\delta}\sum_{k=1}^6|\phi\rangle_{\rm s}^{k}\otimes|\psi\rangle^k_{\rm an} ,
\end{eqnarray}
where the signal mode part $|\phi\rangle_{\rm s}^{k}$ encodes the corresponding Ising ground state, while the ancillary part $|\psi\rangle^k_{\rm an}$ indicates the unsatisfied coupling.

The high fidelity ($\approx0.95$) with respect to the pure dark state $(\ref{NMdarkstate})$ in Fig.~\ref{fig2}(a) implies that the optical effective spin system is not frustrated and resides in a loss-free ground mode.

Next, we discuss the realization of ancillary modes in general cases of multiple coupling flips and the advantages in the semiclassical regime. Consider the following coupling terms,
\begin{eqnarray}\label{fegeneral}
L_{\rm ans}^{n,m}&=&L_{n,m}+i(b_{\rm control}+b_{n,m})+2a_{\rm ans}^{n,m},\nonumber\\
L_{\rm ani}^{n,m}&=&L_{n,m}-i(b_{\rm control}+b_{n,m})+2a_{\rm ani}^{n,m},\nonumber\\
L_{\rm control}&=&\sum b_{n,m}+(2N_{\rm F}-N_{\rm c})b_{\rm control}.
\end{eqnarray}
Here, $b_{\rm control}$ and $b_{n,m}$ are DOPO ancillary modes, $a_{\rm ans}^{n,m}$ and $a_{\rm ani}^{n,m}$ are ancillary NDOPO modes, and $N_{\rm F}$ is the number of coupling terms flipped. The first two lines in Eq.~(\ref{fegeneral}) imply that a flipped coupling term corresponds to an ancilla with phase $0$, so that the corresponding DOPO ancillary mode $b_{n,m}$ forms a dark mode with the control mode $b_{\rm control}$. The third line in Eq.~(\ref{fegeneral}) constrains the number of $b_{n,m}$ with negative phases according to $b_{\rm control}$ to be $N_{\rm flip}$. Note that Eq.~(\ref{collective loss with pair ancillary}) is not a special case of Eq.~(\ref{fegeneral}). The scheme presented in~(\ref{collective loss with pair ancillary}) is more resource efficient but not suitable for flipping multiple couplings.

We apply Eq.~(\ref{fegeneral}) to the model in Fig.~\ref{fig1}(a) and study its advantages in addition to quantum effects by simulating the corresponding semiclassical equations~\cite{SupplementalMaterial}. The relation between the logarithm of the amplitude inhomogeneity and the Ising energy of the solution candidates is calculated with $16^4$ random initial conditions and shown in Fig.~\ref{fig2}(b). The results with $N_{\rm F}=0,1$ reflect the frustration elimination process illustrated in Fig.~\ref{fig1}(d). Amplitude inhomogeneity is large due to frustration for $N_{\rm F}=0$, and can be strongly suppressed for $N_{\rm F}=1$. Although we may get wrong solution candidates, these errors can be identified by the amplitude inhomogeneity. In addition to the error detection, the frustration elimination method can also be used to search for excited states, as shown with the results for $N_{\rm F}=2,3$. By flipping more coupling terms $N_{\rm F}=3$, the global minimum with strongly suppressed amplitude inhomogeneity captures the degenerated first excited state. More examples are provided in the Supplemental Materials~\cite{SupplementalMaterial}.

{\it Realization of the frustration-elimination setup---}The NDOPO modes required in Eq.~(\ref{collective loss with pair ancillary}) can be realized with an additional fiber loop~\cite{10.1364/OPTICA.6.001361, 10.1364/OPTICA.415569, 10.1038/s41467-021-21048-z}, and the collective loss terms in Eq.~(\ref{collective loss with pair ancillary}) may be realized by delay lines with multiple ports~\cite{SupplementalMaterial}, or feedback based on a field-programmable gate array (FPGA). According to Eq.~(\ref{fegeneral}), the necessary number of ancillary modes for a Ising model with $N_{\rm c}$ coupling terms is $(3N_{\rm c}+1)$.

The proposed method can also be extended to qubit-based Ising machines~\cite{10.1038/nature10012} leveraging the strong nonlinearity in superconducting circuits~\cite{10.1103/PhysRevLett.129.220501,10.1103/PhysRevLett.132.240601}, e.g., in a coupling $J_{n,m}\sigma_{z}^{(n)}\sigma_{z}^{(m)}\sigma_{z,\rm an}^{(n,m)}$. The total number of flipped coupling terms can be constrained by a coupling among the ancillary qubits $H_{\rm I}=\sum g\sigma^{(n,m)}_{-,\rm an}\sigma^{(n',m')}_{+,\rm an}+{\rm H.c.}$, which conserves the total number of ancillary qubits with up direction. This extended method could also be applied to some optical Ising machines with strong nonlinearity~\cite{10.1038/s42005-020-0376-5}. In the semiclassical regime, our scheme can also be applied in oscillator-based Ising machines~\cite{10.1038/ncomms15785,10.1038/srep21686} and spatial CIMs~\cite{10.1103/PhysRevApplied.16.054022}, as the coherent couplings applied in these systems are equivalent to the feedback pulses calculated by a field-programmable gate array (FPGA).

{\it Conclusions---}We have developed a method  for mapping frustrated Ising models to optical effective spin systems without frustration, while preserving the set of ground states. In our proposal, the DOPO modes perform, similar to current CIMs, the ground state search, while additional ancillary modes autonomously identify and flip the coupling terms causing frustration. The ancillary modes and switchable couplings can be realized by NDOPO modes and multiport optical couplings, respectively. As DOPOs and NDOPOs can share the same pump field, our modified CIM can be operated in line with the current protocol, i.e., by gradually increasing the pump intensity. Moreover, our numerical results confirm that the system produces a superposition of all the desired solutions when following the common CIM solution-searching protocol. Although researchers debate on whether CIMs with mixed final state benefit from quantum effects, the quantum advantages of pure final states enabled by our frustration-elimination method have been widely studied in spin systems~\cite{SupplementalMaterial,10.1016/j.physrep.2011.08.003,10.1038/nature16176,10.1103/RevModPhys.90.035005,10.1103/PhysRevA.97.042112,10.1038/s41467-020-19495-1}. For CIMs operated in the semiclassical regime, our method can realize self-error detection and excited-state searching.

Our proposal makes it possible to avoid the intrinsic single-photon loss caused by frustration in the current CIM design, so that the exploitation of loss-vulnerable quantum effects becomes conceivable. Such quantum effects would not only be essential for developing quantum CIMs, but may also help in the solution searching. In addition, other severe problems related to frustration (in particular, detrimental amplitude inhomogeneity) are resolved by frustration elimination. Finally, our proposal is also compatible with the measurement-feedback coupling paradigm, and there, too, can enable error detection and excited-state searching.
%
%


\begin{acknowledgments}
{\it Acknowledgments---}J.Q.Y. is partially supported by the National Key Research and Development Program of China (Grant No.~2022YFA1405200) and the National Natural Science Foundation of China (NSFC) (Grant No.~92265202 and No.~11934010). C.G. is partially supported by RIKEN Incentive Research Projects. F.N. is supported in part by: Nippon Telegraph and Telephone Corporation (NTT) Research, the Japan Science and Technology Agency (JST) [via the CREST Quantum Frontiers program Grant No. JPMJCR24I2, the Quantum Leap Flagship Program (Q-LEAP), and the Moonshot R{\&}D Grant Number JPMJMS2061], and the Office of Naval Research (ONR) Global (via Grant No. N62909-23-1-2074). Z.Y.Z. is partially supported by the ZSTU intramural grant (23062089-Y) and the National Natural Science Foundation of China (Grant No.~12371135).
\end{acknowledgments}

\end{document}


\title{Supplemental Material for:\\
Frustration elimination and excited state search in coherent Ising machines}

\author{Zheng-Yang Zhou}
\affiliation{Zhejiang Key Laboratory of Quantum State Control and Optical Field Manipulation, Department of Physics, Zhejiang Sci-Tech University, Hangzhou 310018, China}
\affiliation{\mbox{Theoretical Quantum Physics Laboratory, Cluster for Pioneering Research,} RIKEN, Wakoshi, Saitama 351-0198, Japan}
\author{Clemens Gneiting}
\affiliation{\mbox{Theoretical Quantum Physics Laboratory, Cluster for Pioneering Research,} RIKEN, Wakoshi, Saitama 351-0198, Japan}
\affiliation{\mbox{Center for Quantum Computing, RIKEN, Wakoshi, Saitama 351-0198, Japan}}
\author{J. Q. You}
\altaffiliation[jqyou@zju.edu.cn]{}
\affiliation{Zhejiang Key Laboratory of Micro-Nano Quantum Chips and Quantum Control, School of Physics, School of Physics, and State Key Laboratory for Extreme Photonics and Instrumentation, Zhejiang University, Hangzhou 310027, China}
\affiliation{College of Optical Science and Engineering, Zhejiang University, Hangzhou 310027, China}
\author{Franco Nori}
\altaffiliation[fnori@riken.jp]{}
\affiliation{\mbox{Theoretical Quantum Physics Laboratory, Cluster for Pioneering Research,} RIKEN, Wakoshi, Saitama 351-0198, Japan}
\affiliation{\mbox{Center for Quantum Computing, RIKEN, Wakoshi, Saitama 351-0198, Japan}}
\affiliation{\mbox{Physics Department, The University of Michigan, Ann Arbor, Michigan 48109-1040, USA}}
\maketitle
\renewcommand{\theequation}{S\arabic{equation}}
\renewcommand{\thefigure}{S\arabic{figure}}
\section{Mapping Ising models to Coherent Ising Machines}
The basic structure of a CIM is a fiber cavity complemented by a nonlinear crystal and a coupling module. Optical pulses propagate in the cavity and form degenerated optical parametric oscillators (DOPOs) described by a two-photon pump (in the interaction picture),
\begin{eqnarray}\label{twophotonpump}
H=-iS\sum_n[(a_{n}^{\dag})^2-(a_{n})^2],
\end{eqnarray}
and loss terms including two-photon loss,
\begin{eqnarray}\label{twophotoloss}
\mathcal{L}_{\rm tp}(\rho)&=&\sum_n\frac{\Gamma_{\rm tp}}{2}[2a_{n}a_{n}\rho(t)a_{n}^{\dag}a_{n}^{\dag}-\{a_{n}^{\dag}a_{n}^{\dag}a_{n}a_{n},\rho(t)\}],\nonumber\\
\end{eqnarray}
and single-photon loss,
\begin{eqnarray}\label{singlephotoloss}
\mathcal{L}_{\rm s}(\rho)&=&\sum_n\frac{\Gamma_{\rm s}}{2}[2a_n\rho(t)a_n^{\dag}-\{a_n^{\dag}a_n,\rho(t)\}],\nonumber\\
\end{eqnarray}
where $a_n$ is the annihilation operator of the $n$th DOPO mode, $\{\bullet,\bullet\}$ denotes the anti-commutator, and $\rho$ is the density matrix describing all the DOPO modes. The DOPO exhibits a phase transition at
\begin{eqnarray}\label{qptp}
2|S|=\Gamma_{s},
\end{eqnarray}
where the steady-state transitions from a squeezed vacuum state to two possible coherent states~\cite{10.1103/PhysRevLett.60.1836,10.1103/PhysRevLett.95.083601}:
\begin{eqnarray}
|\Psi(t\rightarrow\infty)\rangle=|\pm\alpha\rangle.
\end{eqnarray}
CIMs use these two coherent states to emulate spin states:
\begin{eqnarray}
|\alpha\rangle\longleftrightarrow|\uparrow\rangle,~~|-\alpha\rangle\longleftrightarrow|\downarrow\rangle.
\end{eqnarray}
If there are $N$ DOPO pulses in the CIM, the steady state is a collective mode corresponding to a many-body spin system, e.g.,
$$|\alpha\rangle\otimes|-\alpha\rangle\dots\otimes|\alpha\rangle\longleftrightarrow|\uparrow\rangle\otimes|\downarrow\rangle\dots\otimes|\uparrow\rangle.$$

Note that a DOPO exhibits dark states if the single-photon loss rate is vanishing $(\Gamma_{\rm s}\approx0)$:
\begin{eqnarray}{\label{darkspace}}
|\Psi(t\rightarrow\infty)\rangle=C_{+}|\alpha\rangle+C_{-}|-\alpha\rangle,
\end{eqnarray}
and the complex amplitude $\alpha$ of the coherent states is given by $\alpha=i\sqrt{2S/\Gamma_{\rm d}}$.

Two common design principles for implementing the optical coupling in CIMs are the optical delay-line architecture~\cite{10.1038/nphoton.2014.249} and the measurement-feedback architecture~\cite{10.1126/science.aah5178}, respectively. The optical delay-line coupling between two DOPO modes can be described by a collective loss,
\begin{eqnarray}\label{optical Ising couplingsm}
\mathcal{L}_{m,n}(\rho)&=&\frac{\Gamma_{\rm c}}{2}(2L_{n,m}\rho L^{\dag}_{n,m}-L^{\dag}_{n,m}L_{n,m}\rho-\rho L^{\dag}_{n,m}L_{n,m}),\nonumber\\
L_{n,m}&=&a_n+{\rm sign}(J_{m,n})a_m.
\end{eqnarray}
where $n$ and $m$ correspond to two different DOPO modes. This phase-dependent loss in Eq.~(\ref{optical Ising couplingsm}) can be associated with an Ising interaction term:
\begin{eqnarray}
H_{n,m}=J_{m,n}\sigma_z^{(n)}\sigma_{z}^{(m)},
\end{eqnarray}
where $\sigma_z^{(n)}$ is the Pauli matrix of the $n$th spin, and the coupling strength satisfies $|J_{m,n}|=J$. The effect of the collective loss and of the Ising interaction, respectively, are summarized in the following table:\\

\begin{tabular}{|c|c|c|c|}
  \hline
  ~ & Effect& States &Values\\
  \hline
  {\rm Collective loss}& Photon loss probability & $|\alpha\rangle_n\otimes|\alpha\rangle_m$ &$2|\alpha_n|^2\Gamma_{\rm c}(1+{\rm sign}(J_{m,n}))$\\
  ~& ~& $|\alpha\rangle_n\otimes|-\alpha\rangle_m$&$2|\alpha_n|^2\Gamma_{\rm c}(1-{\rm sign}(J_{m,n}))$\\
  ~&~&$|-\alpha\rangle_n\otimes|-\alpha\rangle_m$ & $2|\alpha_n|^2\Gamma_{\rm c}(1+{\rm sign}(J_{m,n}))$\\
  ~&~&$|-\alpha\rangle_n\otimes|\alpha\rangle_m$ &$2|\alpha_n|^2\Gamma_{\rm c}(1-{\rm sign}(J_{m,n}))$\\
  \hline
  {\rm Ising interaction} &Energy shift& $|\uparrow\rangle_n\otimes|\uparrow\rangle_m$&$J_{m,n}$\\
  ~&~& $|\uparrow\rangle_n\otimes|\downarrow\rangle_m$&$-J_{m,n}$\\
  ~&~ &$|\downarrow\rangle_n\otimes|\downarrow\rangle_m$ &$J_{m,n}$\\
  ~&~&$|\downarrow\rangle_n\otimes|\uparrow\rangle_m$ &$-J_{m,n}$\\
  \hline
\end{tabular}\\\\
It is straightforward to see that the energy difference in the spin system is mapped to the loss difference in DOPO modes. An Ising energy $-J$ corresponds to vanishing loss, while an Ising energy $J$ corresponds to finite loss. Due to the loss-dependent phase transition in Eq.~(\ref{qptp}), the Ising ground state is mapped to a collective DOPO mode with lowest transition pump strength $|S|$.

The measurement-feedback coupling can be expressed as classical pumps on different DOPO modes:
\begin{eqnarray}\label{mfc}
H_{\rm MF}&=&-i\sum_{n,m}\Omega ~{\rm sign}(J_{n,m})\langle(a_m+a_m^\dag)\rangle(a_n-a_n^\dag).
\end{eqnarray}
Such a pump can effectively modify the two-photon pump strength $S$ in the semiclassical limit (mean field):
\begin{eqnarray}\label{sm:measurementfeedback}
H+H_{\rm MF}&=&-iS\sum_n[(a_{n}^{\dag})^2-(a_{n})^2]-i\sum_{n,m}\Omega ~{\rm sign}(J_{n,m})\langle(a_m+a_m^\dag)\rangle(a_n-a_n^\dag)\nonumber\\
            &\approx&-i\sum_n[S\langle(a_n+a_n^\dag)\rangle-\sum_m\Omega~ {\rm sign}(J_{n,m})\langle(a_m+a_m^\dag)\rangle](a^\dag_n-a_n).
\end{eqnarray}
As the steady states of DOPOs are coherent states with $0$ phase or $\pi$ phase, we have
$$\langle(a_n+a_n^\dag)\rangle=\pm\langle(a_m+a_m^\dag)\rangle,$$
in the semi-classical limit. Here, we assume a negative $S$ to make the amplitude $\alpha$ real. If
$$\langle(a_n+a_n^\dag)\rangle={\rm sign}(J_{n,m})\langle(a_m+a_m^\dag)\rangle,$$
the total pump strength acting on the $n$th mode is increased. For an opposite relative phase
$$\langle(a_n+a_n^\dag)\rangle=-{\rm sign}(J_{n,m})\langle(a_m+a_m^\dag)\rangle,$$
the total pump strength acting on the $n$th mode is reduced. According to the threshold relation in Eq.~(\ref{qptp}), modifying the pump strength of the collective mode is equivalent to modifying the loss of the collective mode. Therefore, the measurement-feedback in Eq.~(\ref{sm:measurementfeedback}) can also be mapped to the Ising interaction.

\section{Problems caused by frustration}
\subsection{Intrinsic loss}
In the collective loss coupling protocol, the Ising energy is mapped to the loss. Note that a collective loss with zero loss (that is, the system is in a dark state) corresponds to a spin configuration in which all the Ising coupling terms contribute negative energies. Such spin configurations do not exist in frustrated Ising models. Therefore, there will be intrinsic loss caused by the coupling protocol, if we use CIMs to simulate frustrated Ising models. Such intrinsic loss generically destroys most strong quantum effects, and thus prevents CIMs benefiting from quantum effects. Note that the current measurement-feedback coupling protocol exhibits few quantum effects.

\subsection{Inhomogeneity in amplitudes}
Although the current CIMs mainly work in the semi-classical regime, frustration can still cause problems. To correctly map an Ising model to a CIM, the amplitudes of different steady-state DOPO pulses are required to be the same. To see how inhomogeneous amplitudes can cause errors, let us, for example, consider a pair of spins, $$|\uparrow\rangle_n\otimes|\uparrow\rangle_m,$$ which has the Ising coupling energy $-J$ for the coupling term
$$H_{n,m}=-J\sigma_z^n\sigma_z^m.$$
However, the collective DOPO mode
$$|\alpha\rangle_n\otimes|\alpha+\Delta\rangle_m,$$
is not a dark mode of the collective loss operator,
$$L_{n,m}=a_n-a_m.$$
In addition the effective pump strength shift in Eq.~(\ref{sm:measurementfeedback}) requires that
$$|\langle(a_m+a_m^\dag)\rangle|=|\langle(a_n+a_n^\dag)\rangle|.$$

Note that one possible approach to mitigate the problem of inhomogeneous amplitudes is the chaotic amplitude control~\cite{10.1103/PhysRevLett.122.040607}.
\section{Potential advantages of quantum states in a CIM}
The advantages of quantum states for metrology have been analytically proved in spin systems~\cite{10.1016/j.physrep.2011.08.003}. Following the idea of spin systems, we show the potential advantages of quantum final states in CIMs. Consider an Ising problem with $N_{\rm solution}$ degenerate ground states and the corresponding optical states in a CIM
\begin{eqnarray}
|\Psi_n\rangle,~~0\leq n<N_{\rm solution}.
\end{eqnarray}
Due to the presence of $N_{\rm solution}$ degenerate ground states, the probability of obtaining a specific solution is
\begin{eqnarray}
P=\frac{1}{N_{\rm solution}}.
\end{eqnarray}
Therefore, it is hard to verify the correctness of a candidate solution $|\psi_{\rm can}\rangle$ if the CIM produces a mixture of the solution states $|\Psi_n\rangle$. Now we consider a superposition of all the ground states,
\begin{eqnarray}
|\Psi\rangle=\frac{1}{\sqrt{N_{\rm solution}}}\sum_n|\Psi_n\rangle.
\end{eqnarray}
Here, the overlap between different effective spin states in CIMs is assumed to be negligible $\langle\Psi_{n}|\Psi_{m}\rangle\approx 0$ for $n\ne m$. Assume now that we have obtained $N_{\rm det}$ determined ground states $|\Psi_{n,\rm det}\rangle$ with $n<N_{\rm det}$. We can then prepare a superposition of all these determined ground states,
\begin{eqnarray}
|\Psi_{\rm det}\rangle=\frac{1}{\sqrt{N_{\rm det}}}\sum_n|\Psi_{n,\rm det}\rangle,
\end{eqnarray}
and use the coherence between different solutions to verify the correctness of the candidate state
\begin{eqnarray}
P_{\rm co}=\langle \Psi_{\rm det}|\Psi\rangle\langle\Psi|\Psi_{\rm can}\rangle+\langle \Psi_{\rm can}|\Psi\rangle\langle\Psi|\Psi_{\rm det}\rangle=\frac{2\sqrt{N_{\rm det}}}{N_{\rm solution}}.
\end{eqnarray}
We find that the coherence has a higher probability to be measured compared to the projection measurement with the success probability $1/N_{\rm solution}$.
\section{The idea of The ancillary-mode construction}
\begin{figure}[h]
\center
\includegraphics[width=6in]{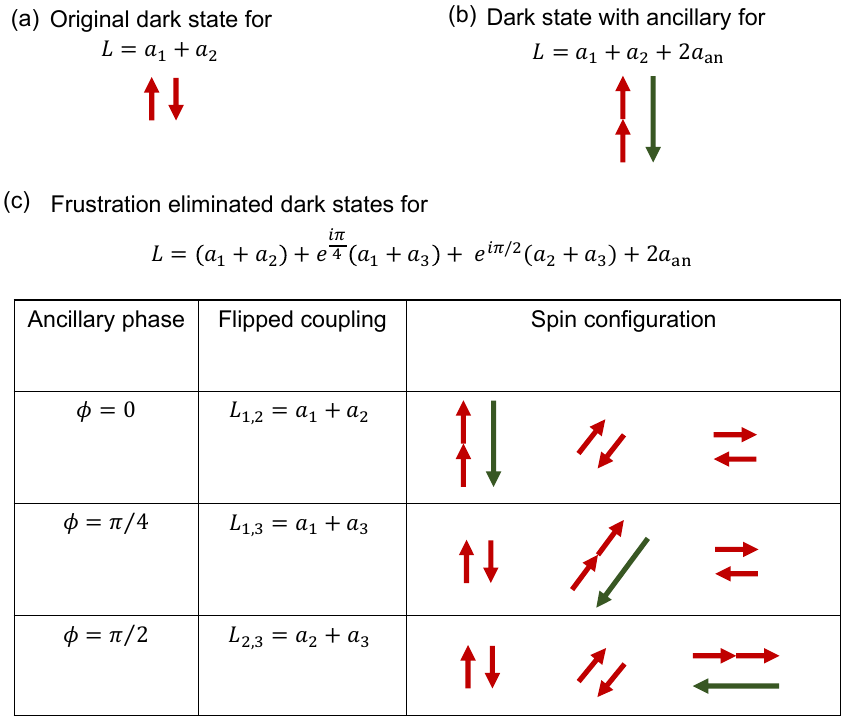}
\caption{Illustration of the ancillary-controlled coupling flipping.}
\label{sm:fig0d5}
\end{figure}
Here we introduce the basic idea behind introducing ancillary modes in dissipative coupling channels. Consider the collective coupling illustrated in Fig.~\ref{sm:fig0d5}(a), which can be expressed by the Lindblad operator,
\begin{eqnarray}
L=a_1+a_2.
\end{eqnarray}
When the two effective spins have opposite orientations, they cancel in the loss channel and form a dark mode, as illustrated in Fig.~\ref{sm:fig0d5}(a). On the other hand, the collective modes associated to aligned orientations couple to the loss channel, so that this alignment is suppressed by the dissipative coupling. This way, the dissipative coupling in Fig.~\ref{sm:fig0d5}(a) induces an antiferromagnetic coupling.

The preferred collective spin state can be modified by adding an additional ancillary mode, as illustrated in Fig.~\ref{sm:fig0d5}(b),
\begin{eqnarray}
L'=a_1+a_2+2a_{\rm an}.
\end{eqnarray}
Due to the presence of a ``larger'' ancillary spin, the two signal spins must have the same orientation for all three spins to form a dark mode. In the new dark mode the two signal spins have the same orientation, as illustrated in Fig.~\ref{sm:fig0d5}(b). Note that the original dark mode $(a_1-a_2)/\sqrt{2}$ is still decoupled from the loss channel in Fig.~\ref{sm:fig0d5}(b), but the original bright mode $(a_1+a_2)/\sqrt{2}$ is now pumped by the ancillary mode $a_{\rm an}$. This can be verified as follows,
\begin{eqnarray}
\frac{\partial \langle a_1+a_2\rangle}{\partial t}&=&-i\langle[H,\rho](a_1+a_2)\rangle+\langle\mathcal{L}_{\rm tp}(\rho)(a_1+a_2)\rangle\nonumber\\
&&+\frac{\Gamma_{\rm c}}{2}\langle (2L'\rho {L'}^{\dag}-{L'}^{\dag}L'\rho-\rho {L'}^{\dag}L')(a_1+a_2)\rangle\nonumber\\
&=&-i\langle[H,\rho](a_1+a_2)\rangle+\langle\mathcal{L}_{\rm tp}(\rho)(a_1+a_2)\rangle\nonumber\\
&&+\frac{\Gamma_{\rm c}}{2}[\langle L'\rho {L'}^{\dag}(a_1+a_2)\rangle-\langle{L'}^{\dag}L'\rho(a_1+a_2)\rangle+\langle L'\rho {L'}^{\dag}(a_1+a_2)\rangle-\langle\rho {L'}^{\dag}L'(a_1+a_2)\rangle]\nonumber\\
&=&-i\langle[H,\rho](a_1+a_2)\rangle+\langle\mathcal{L}_{\rm tp}(\rho)(a_1+a_2)\rangle\nonumber\\
&&+\frac{\Gamma_{\rm c}}{2}[\langle L'\rho {L'}^{\dag}(a_1+a_2)\rangle-\langle L'\rho(a_1+a_2){L'}^{\dag}\rangle+\langle \rho {L'}^{\dag}(a_1+a_2)L'\rangle-\langle\rho {L'}^{\dag}L'(a_1+a_2)\rangle]\nonumber\\
&=&-i\langle[H,\rho](a_1+a_2)\rangle+\langle\mathcal{L}_{\rm tp}(\rho)(a_1+a_2)\rangle\nonumber\\
&&+\frac{\Gamma_{\rm c}}{2}[\langle L'\rho[ {L'}^{\dag},(a_1+a_2)]\rangle+\langle \rho {L'}^{\dag}[(a_1+a_2),L']\rangle]\nonumber\\
&=&-i\langle[H,\rho](a_1+a_2)\rangle+\langle\mathcal{L}_{\rm tp}(\rho)(a_1+a_2)\rangle\nonumber\\
&&+\frac{\Gamma_{\rm c}}{2}\langle (a_1+a_2+2a_{\rm an})\rho[ (a_1^{\dag}+a_2^{\dag}+2a_{\rm an}^{\dag}),(a_1+a_2)]\rangle\nonumber\\
&&+\frac{\Gamma_{\rm c}}{2}\langle \rho (a_1^{\dag}+a_2^{\dag}+2a_{\rm an}^{\dag})[(a_1+a_2),(a_1+a_2+2a_{\rm an})]\rangle\nonumber\\
                                                  &=&-i\langle[H,\rho](a_1+a_2)\rangle+\langle\mathcal{L}_{\rm tp}(\rho)(a_1+a_2)\rangle-\Gamma_{\rm c}(\langle a_1\rangle+\langle a_2\rangle+2\langle a_{\rm an}\rangle).
\end{eqnarray}
Note that the last term $-2\Gamma_{\rm c}\langle a_{\rm an}\rangle$ can be an effective pump if $\langle a_{\rm an}\rangle$ has the same sign as $-(\langle a_1\rangle+\langle a_2\rangle)$. Such an effective pump can cause the mode $(a_1+a_2)$ to dominate the competition with the original dark mode $(a_1-a_2)$, which is induced by the two-photon loss terms. Such a competition mechanism can be understood as follows,
\begin{eqnarray}
\langle\mathcal{L}_{\rm tp}(\rho)(a_1+a_2)\rangle&=&\frac{\Gamma_{\rm tp}}{2}\sum_{k=1,2}\langle a_k^2 \rho [(a^{\dag}_k)^2,(a_1+a_2)]\rangle\nonumber\\
&=&-\Gamma_{\rm tp}\langle  a^{\dag}_1a_1^2\rangle-\Gamma_{\rm tp}\langle  a^{\dag}_2a_2^2\rangle\nonumber\\
&=&-
\frac{\Gamma_{\rm tp}}{2\sqrt{2}}\langle  (a_+^{\dag}+a_-^{\dag})(a_++a_-)^2\rangle-\frac{\Gamma_{\rm tp}}{2\sqrt{2}}\langle  (a^{\dag}_+-a_-^{\dag})(a_+-a_-)^2\rangle\nonumber\\
&=&-
\frac{\Gamma_{\rm tp}}{2\sqrt{2}}\langle  (a_+^{\dag}a_+^2+2a_+^{\dag}a_-a_++a_+^{\dag}a_-^2)\rangle-
\frac{\Gamma_{\rm tp}}{2\sqrt{2}}\langle  (a_-^{\dag}a_+^2+2a_-^{\dag}a_-a_++a_-^{\dag}a_-^2)\rangle\nonumber\\
&&-\frac{\Gamma_{\rm tp}}{2\sqrt{2}}\langle  (a^{\dag}_+a_+^2-2a^{\dag}_+a_-a_++a^{\dag}_+a_-^2)\rangle+\frac{\Gamma_{\rm tp}}{2\sqrt{2}}\langle  (a_-^{\dag}a_+^2-2a_-^{\dag}a_-a_++a_-^{\dag}a_-^2)\rangle\nonumber\\
&=&-\frac{\Gamma_{\rm tp}}{\sqrt{2}}(\langle  a_+^{\dag}a_+^2\rangle+\langle a_+^{\dag}a_-^2\rangle+\langle  2a_-^{\dag}a_-a_+\rangle).
\end{eqnarray}
with $a_+=(a_1+a_2)/\sqrt{2}$ and $a_-=(a_1-a_2)/\sqrt{2}$. Under the mean field approximation, the last two terms describe single-photon loss acting on the mode $a_+$ with a loss rate depending on the semiclassical amplitude of the mode $a_-$. It follows that $a_-$ is suppressed by a pumped $a_+$, even though it is decoupled from the loss.

To control multiple couplings with a single ancillary mode, we may associate different coupling terms with different phases and use a hyperspin as the ancillary mode, as illustrated in Fig.~\ref{sm:fig0d5}(c). The phase is an effective rotation of the $z$ direction of each coupling term, so that different coupling terms do not interfere with each other. The hyperspin ancilla chooses a specific spin orientation, and flips the corresponding coupling term.
\section{Semiclassical equations for a CIM}
In the limit of large amplitudes, we can also approximate a CIM with semiclassical equations of motion, which follow from a mean field approximation of the quantum dynamics:
\begin{eqnarray}\label{sm:CIMsc}
\frac{\partial}{\partial t}A_{n}&=&PA_{n}^*-\gamma_{\rm s}A_{n}-\gamma_{\rm d}A_{n}|A_{n}|^2-C_{n}\nonumber\\
C_n&=&\sum_{m}|J_{n,m}|[A_n+{\rm sign}(J_{n,m})A_m],
\end{eqnarray}
where $A_n\equiv\langle a_n\rangle$ is the amplitude of the $n$th mode with the annihilation operator $a_n$, and $C_n$ is the coupling term determined by the coupling matrix $J_{n,m} = \pm J$ with $J>0$. Without coupling $|J_{n,m}|=0$, each mode in Eq.~(\ref{sm:CIMsc}) has two steady solutions $A_m=\pm\sqrt{(P-\gamma_{\rm s})/\gamma_{\rm d}}$, which correspond to two spin orientations, above the threshold $P>\gamma_{\rm s}$.

The coupling term $C_n$ can shift the threshold as
\begin{eqnarray}
P>\gamma_{\rm s}+\sum_{m}|J_{n,m}|+\sum_{j}J_{n,m}\frac{A_m}{A_n}.
\end{eqnarray}
Note that $A_{m}/A_{n}=\pm 1$, if the amplitudes are homogeneous.
\subsection{Semiclassical description for frustration elimination}
The ancillary modes in the frustration-elimination method can be described by the following mean-field equations,
\begin{eqnarray}\label{sm:scNDOPO}
\frac{\partial}{\partial t}A_{\rm ans}&=&2SA_{\rm ani}^*-\Gamma_{\rm tp}A_{\rm ans}|A_{\rm ani}|^2-C_{\rm ans},\nonumber\\
\frac{\partial}{\partial t}A_{\rm ani}&=&2SA_{\rm ans}^*-\Gamma_{\rm tp}A_{\rm ani}|A_{\rm ans}|^2-C_{\rm ani},
\end{eqnarray}
with $A_{\rm ans}\equiv\langle a_{\rm ans}\rangle$ and $A_{\rm ani}\equiv\langle a_{\rm ani}\rangle$. Note that the single-photon loss is omitted here. The dark mode condition for the ancillary modes is $A_{\rm ans}A_{\rm ani}=2S/\Gamma_{\rm tp}$. With a pair of conjugate coupling terms $C_{\rm ans}=C_{\rm ani}^*$, the solutions obey the symmetry $A_{\rm ans}=A_{\rm ani}^*$, which is the condition required for the ancillary modes. Note that the Lindblad terms in the main text satisfy this relation.

The semiclassical coupling terms for the frustration elimination are
\begin{eqnarray}\label{feclassicalcoupling}
C_{i}^{\rm F}&=&J^{\rm eff}_i(\tilde{A}+2A_{\rm ans})+{J^{\rm eff}_i}^*(\tilde{A}^*+2A_{\rm ani}),\nonumber\\
C_{\rm ans}&=&2J\{\tilde{A}+2A_{\rm ans}\},~~~~C_{\rm ani}=2J\{\tilde{A}^*+2A_{\rm ani}\},
\end{eqnarray}
with $\tilde{A}=\sum_{n,m}e^{i\phi_{n,m}}[A_m+A_{n}]$, and the phase terms $\phi_{1,2}=0$, $\phi_{2,3}=\pi/2$, $\phi_{1,3}=\pi/4$. Note that the effective coupling strength $J^{\rm eff}_i\equiv \sum_{j}(e^{-i\phi_{i,j}}J_{i,j}+e^{-i\phi_{j,i}}|J_{j,i}|)$, which has no effect on the dark-state condition $C_{i}^{\rm F}=0$, originates from the quantum coupling model described by the Lindblad operator.
\section{Realization of frustration-elimination-type collective loss}
\subsection{Two-mode collective loss with optical delay lines}
\begin{figure}[h]
\center
\includegraphics[width=6.5in]{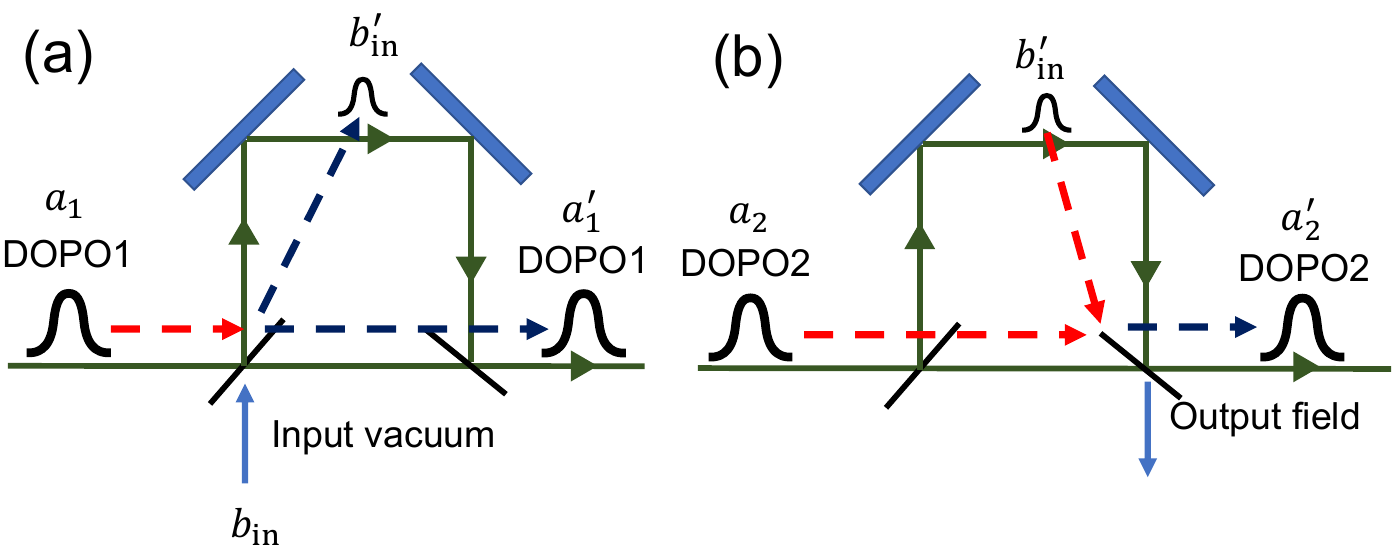}
\caption{Illustration of the collective loss realized through a delay-line coupling. (a) First, the first pulse denoted as $a_1$ passes the first beam splitter and mixes with a vacuum field denoted as $b_{\rm in}$ in the delay line denoted as $b'_{\rm in}$. (b) Then the pulse $b'_{\rm in}$, which carries the information of $a_1$, interacts with the second pulse $a_2$ at the second beam splitter. For a proper collective state, the output field is almost a vacuum state, which corresponds to a dark state of such collective loss.}
\label{sm:fig1}
\end{figure}

We first consider a two-mode collective loss formed by a delay line coupled to two optical pulses, as illustrated in Fig.~\ref{sm:fig1},

The input mode $b_{\rm in}$ (usually the vacuum state) couples with two system modes successively through two beam splitters. The fields after the scattering in Fig.~\ref{sm:fig1}(a) are:
\begin{eqnarray}
b'_{\rm in}&=&Tb_{\rm in}+Ra_1,\nonumber\\
a'_1&=&Ta_1-Rb_{\rm in}.
\end{eqnarray}
After the second scattering, the fields are:
\begin{eqnarray}
b''_{\rm in}&=&T^2b_{\rm in}+RTa_1+Ra_2,\nonumber\\
a'_1&=&Ta_1-Rb_{\rm in},\nonumber\\
a'_2&=&Ta_2-RTb_{\rm in}-R^2a_1.
\end{eqnarray}
Here, the transmission rate is $T^2$ and the reflection rate is $R^2$. When the transmission rate is close to $1$, the input mode $b_{\rm in}$ is not coupled to the collective mode $(a_1-a_2)$ up to the second order of $|R|$. The change of the ``dark mode" is as follows:
\begin{eqnarray}\label{singledelayline}
a'_1-a'_2&=&T(a_1-a_2)+(1-T)Rb_{\rm in}+R^2a_1\nonumber\\
               &=&(T+0.5R^2)(a_1-a_2)+(1-T)Rb_{\rm in}+0.5R^2(a_1+a_2).
\end{eqnarray}
It is easy to see that this ``dark mode" is not really dark, but coupled to the bright mode $(a_1+a_2)$ up to second order in $R$. Therefore, we can conclude that the ``dark mode" in Fig.~\ref{sm:fig1}(a) is dark to the input field $b_{\rm in}$, but not completely decoupled from other modes. Note that the second order of $R$ is important because the loss generated by the delay line on the bright mode is of the order of $R^2$.

\begin{figure}[h]
\center
\includegraphics[width=6.5in]{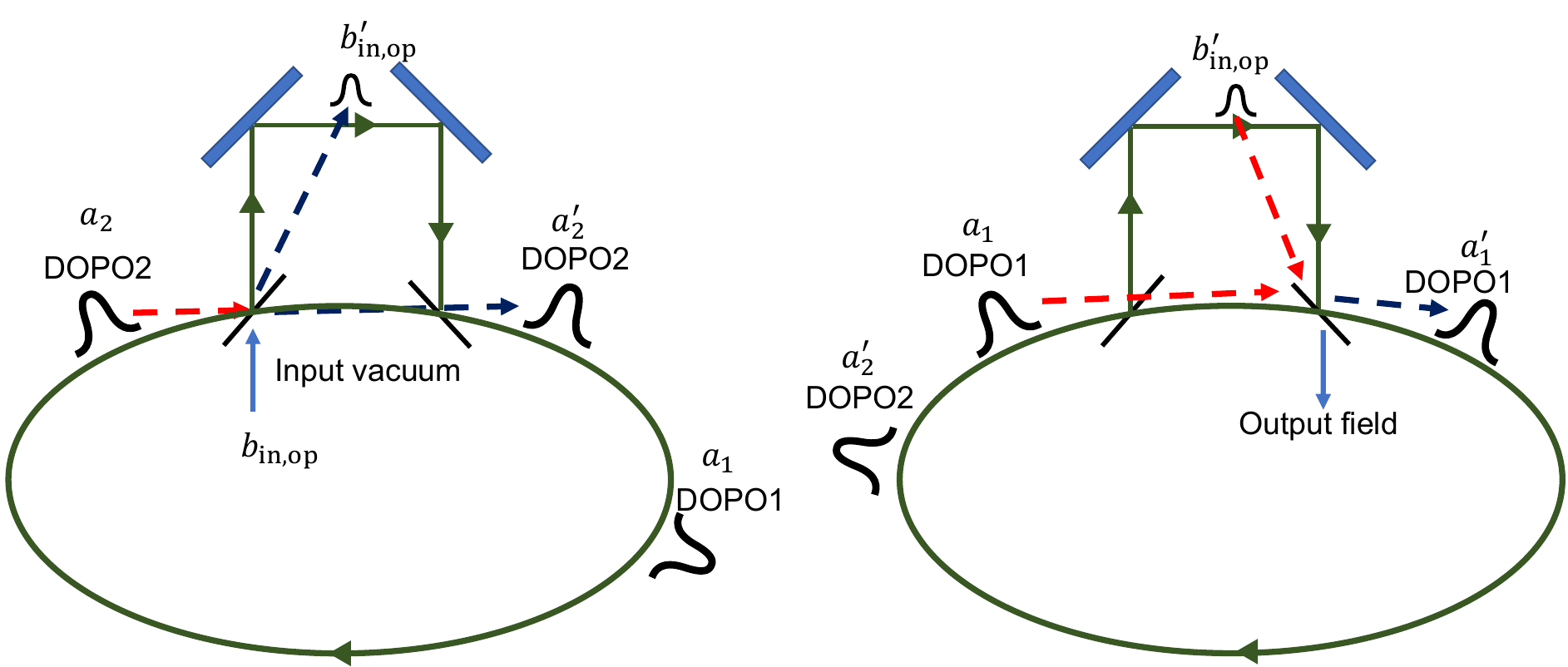}
\caption{Illustration of the open-port delay line in the opposite direction. A delay line can only delay and convey the information of a pulse to the pulses coming after it. However, note that in cyclic structures, e.g., in CIMs, the $n$th pulse in the $m$th circle passes the coupling modula before the $(n-1)$th pulse in the $(m+1)$th circle. Therefore, a delay line in the opposite direction delays the pulses longer than the circling period.}
\label{sm:fig2}
\end{figure}

To solve this problem, we can introduce a delay line with opposite scattering order, as shown in Fig.~\ref{sm:fig2}.
Such a delay line has the following effects on the DOPO pulses:
\begin{eqnarray}
a''_2&=&Ta'_2-Rb_{\rm in,op},\nonumber\\
a''_1&=&Ta'_1-RTb_{\rm in,op}-R^2a'_2.
\end{eqnarray}
The total effects of the two delay lines are:
\begin{eqnarray}
a''_1&=&T^2a_1-T^3Rb_{\rm in}-RTb_{\rm in,op}-R^2Ta_2+R^4a_1,\nonumber\\
a''_2&=&T^2a_2-T^2Rb_{\rm in}-TR^2a_1-Rb_{\rm in,op}.
\end{eqnarray}
After passing the two-delay lines with opposite scattering order, the collective mode $(a_1-a_2)$ becomes,
\begin{eqnarray}
a''_1-a''_2&=&T(T+R^2)(a_1-a_2)-RT^2(T-1)b_{\rm in}-R(T-1)b_{\rm in,op}+R^4a_1\nonumber\\
           &\approx&(a_1-a_2)+o(R^2).
\end{eqnarray}
This collective mode is unchanged up to second order in $T$. Therefore, two delay lines with different directions can form a dark mode. The collective mode with opposite relative phase $(a_1+a_2)$ experiences loss through these delay lines:
\begin{eqnarray}\label{sm:collective loss Langevin}
a''_1+a''_2&=&T(T-R^2)(a_1+a_2)-RT^2(T+1)b_{\rm in}-R(T+1)b_{\rm in,op}+R^4a_1\nonumber\\
           &\approx&(1-R^2)(a_1+a_2)-R(b_{\rm in}+b_{\rm in,op})+o(R^2).
\end{eqnarray}
Note that Eq.~(\ref{sm:collective loss Langevin}) is equivalent to the coupling between a loss channel $b_{\rm in}+b_{\rm in,op}$ and a collective mode $(a_1+a_2)$. The collective loss Eq.~(\ref{sm:collective loss Langevin}) can also be expressed by the Lindblad terms:
\begin{eqnarray}\label{sm:collective loss Lindblad}
\mathcal{L}_{1,2}(\rho)&=&\frac{\Gamma_c}{2}(2L_{1,2}\rho L^{\dag}_{1,2}-L^{\dag}_{1,2}L_{1,2}\rho-\rho L^{\dag}_{1,2}L_{1,2}),\nonumber\\
L_{1,2}&=&a_1+a_2.
\end{eqnarray}
Note that the preferred collective mode can be changed by including an electro-optic phase and amplitude modulator (EOM),
\begin{eqnarray}\label{sm:collective loss change}
L_{\rm 1,2}=a_1+a_2~\longrightarrow~L'_{\rm 1,2}=a_1-a_2.
\end{eqnarray}
\begin{figure}[t]
\center
\includegraphics[width=5in]{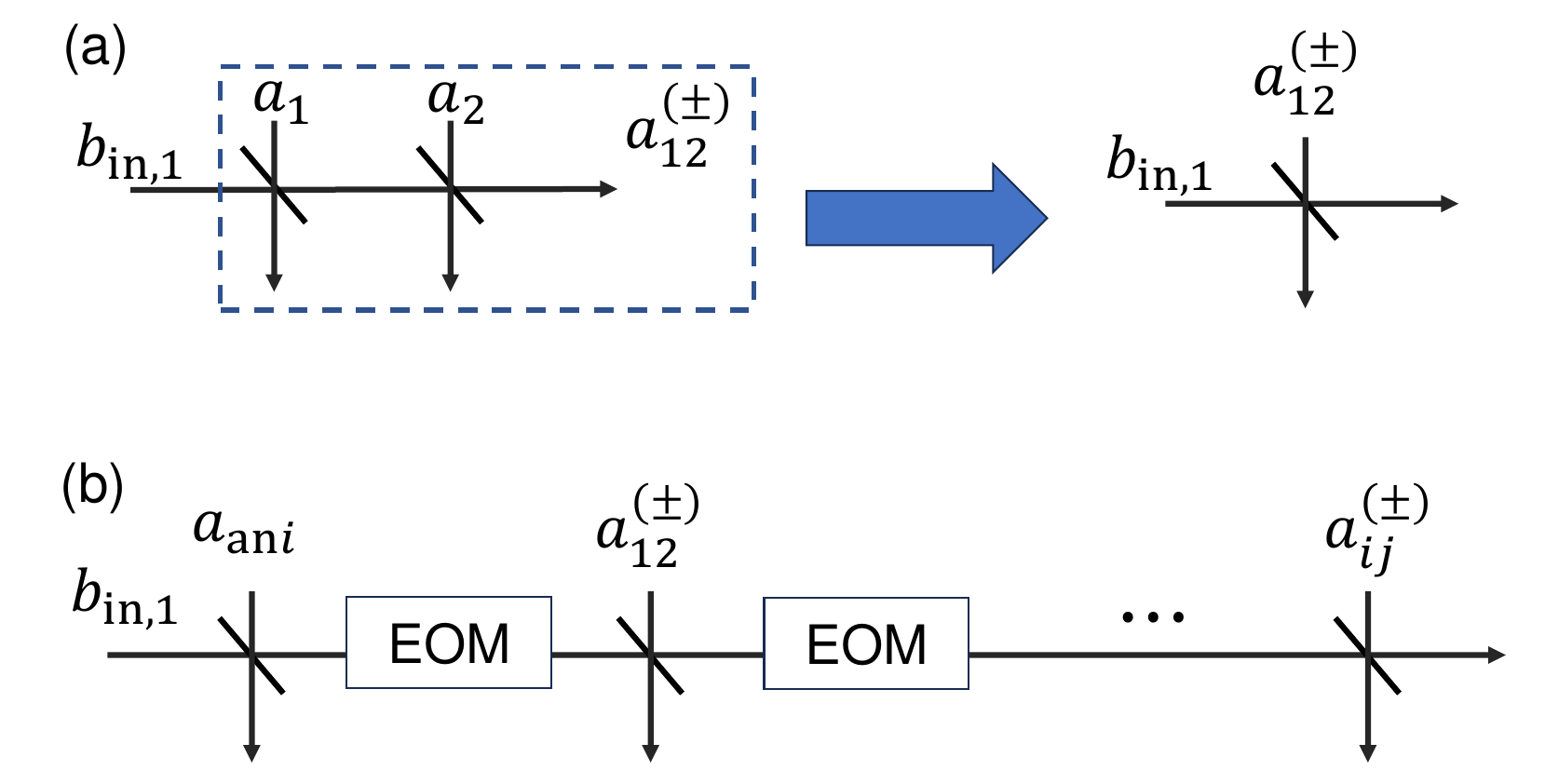}
\caption{Illustration of the multi-port delay line. (a) Expression of two modes coupled to a delay line as a collective mode. (b) Illustrations of phase terms in the frustration-eliminating channel generated by the phase and amplitude modulators (EOM). A multi-port delay line can be naively interpreted as connecting several two-port delay lines with EOMs.}
\label{sm:fig3}
\end{figure}
\subsection{Multi-mode collective loss for frustration elimination}
To eliminate the frustration in CIMs, we need a frustration-eliminating loss channel of the following form:
\begin{eqnarray}\label{sm:multi-mode collective loss}
L_{\rm ani}=\sum_{n,m}e^{i\phi_{m,n}}(a_n+{\rm sign}(J_{m,n})a_m)+2a_{\rm ani}.
\end{eqnarray}
Note that here we take the frustration-eliminating channel of an idler ancillary mode as an example, while the channel for the signal ancillary modes only differs in the phases. Such loss can be realized by multi-port delay lines, as illustrated in Fig.~\ref{sm:fig3}.

\begin{figure}[h]
\center
\includegraphics[width=5in]{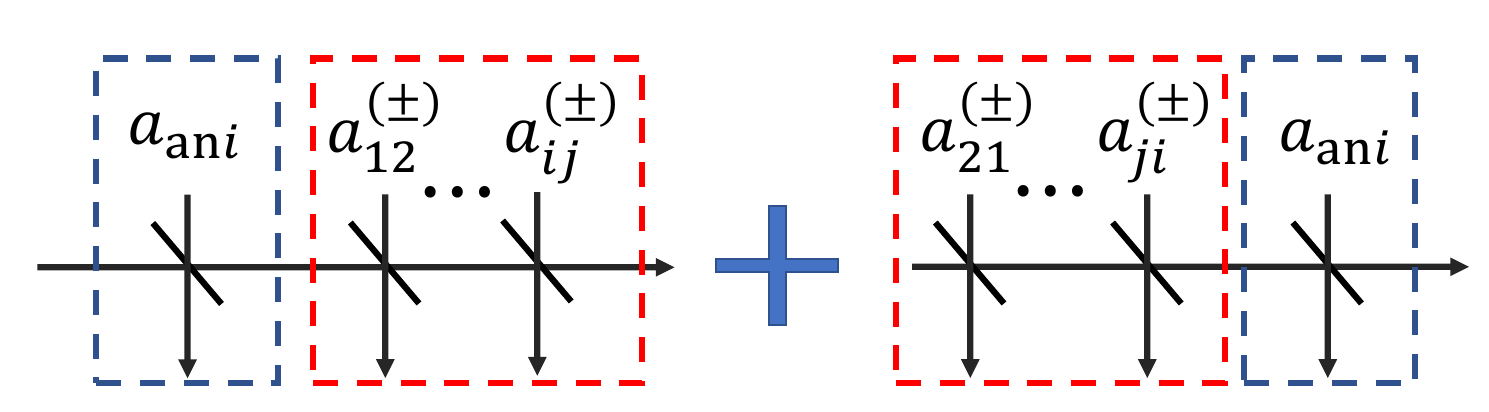}
\caption{Illustration of the frustration-eliminating loss channel formed by a pair of multi-port delay lines. As in the two-port delay cases, the unwanted coupling terms can be eliminated with a delay line in opposite direction.}
\label{sm:fig4}
\end{figure}

To simplify the expressions, we express the two signal modes coupled to the delay line in Fig.~\ref{sm:fig1} as a collective mode, as shown in Fig.~\ref{sm:fig3}(a). Note that
\begin{eqnarray}
a^{(\pm)}_{12}=a_1\pm a_2.
\end{eqnarray}
We can couple multiple collective modes, which correspond to different coupling terms in Eq.~(\ref{optical Ising couplingsm}), to a common open-port delay, as illustrated in Fig.~\ref{sm:fig3}(b). By introducing necessary phase terms $\phi_{m,n}$ with EOMs, the collective mode coupled to the delay line is exactly the Lindblad operator in Eq.~(\ref{sm:multi-mode collective loss}). The additional coupling terms can be cancelled by a delay line in opposite direction as illustrated in Fig.~\ref{sm:fig4}.
\section{Frustration elimination in measurement-feedback coupling}
Following the idea in Eq.~(\ref{sm:measurementfeedback}), the frustration-elimination coupling can also be realized with measurement feedback.
\begin{eqnarray}
H_{\rm FEMF}=i\Omega\langle(L_{\rm ani}+L_{\rm ani}^{\dag})\rangle(L_{\rm ani}-L_{\rm ani}^{\dag}).
\end{eqnarray}
Consider now the semi-classical amplitude equation of the $n$th DOPO mode under the influence of this Hamiltonian,
\begin{eqnarray}
\frac{\partial}{\partial t}\langle a_n\rangle&=&i\langle[H_{\rm FEMF},a_n] \rangle,\nonumber\\
                                             &=&-\Omega\langle(L_{\rm ani}+L_{\rm ani}^{\dag})\rangle\langle[(L_{\rm ani}-L_{\rm ani}^{\dag}),a_n]\rangle
\end{eqnarray}
Note that the contribution of the frustration-elimination loss in Eq.~(\ref{sm:multi-mode collective loss}) to the amplitude equation has a similar form in the semi-classical limit,
\begin{eqnarray}
\frac{\partial}{\partial t}\langle a_n\rangle&=&\frac{\Gamma_{\rm c}}{2}\langle a_n (2L_{\rm ani}\rho L_{\rm ani}^{\dag}-L_{\rm ani}^{\dag}L_{\rm ani}\rho-\rho L_{\rm ani}^{\dag}L_{\rm ani})\rangle,\nonumber\\
                                             &=&\frac{\Gamma_{\rm c}}{2}\langle [a_n,L_{\rm ani}]\rho L_{\rm ani}^{\dag}\rangle+\frac{\Gamma_{\rm c}}{2}\langle [L_{\rm ani}^{\dag},a_n]L_{\rm ani}\rho\rangle,\nonumber\\
                                             &\approx&-\frac{\Gamma_{\rm c}}{2}\langle L_{\rm ani}+L^{\dag}_{\rm ani}\rangle\langle[(L_{\rm ani}-L^{\dag}_{\rm ani}),a_n]\rangle.
\end{eqnarray}\\\\
\section{Scaling of frustration elimination dissipative coupling}
\subsection{General design for scaling the frustration-elimination method}
As mentioned in the main text, applying the frustration elimination method to the general case of flipping several couplings requires a modified model,
\begin{eqnarray}\label{sm:fegeneral}
L_{\rm ans}^{n,m}&=&L_{n,m}+i(b_{\rm control}+b_{n,m})+2a_{\rm ans}^{n,m},\nonumber\\
L_{\rm ani}^{n,m}&=&L_{n,m}-i(b_{\rm control}+b_{n,m})+2a_{\rm ani}^{n,m},\nonumber\\
L_{\rm control}&=&\sum b_{n,m}+(2N_{\rm F}-N_{\rm c})b_{\rm control}.
\end{eqnarray}
Unlike the single-ancilla case, now a pair of frustration eliminating loss channels is attributed to each coupling term $L_{m,n}$. Each pair of ancillary modes can either flip the corresponding loss term or the coupling between a pair of control modes $b_{\rm control}$ and $b_{n,m}$. If a coupling term $L_{n,m}$ is flipped, the corresponding control modes prefer the dark state of $(b_{\rm control}+b_{n,m})$. If we want to flip $N_{\rm F}$ couplings within all the $N_{\rm c}$ coupling terms, we need $N_{\rm F}$ control modes $b_{n,m}$ in the dark state of $(b_{\rm control}+b_{n,m})$ and $(N_{\rm c}-N_{\rm F})$ control modes in the dark state of $(b_{\rm control}-b_{n,m})$. Such control is achieved by the third line of Eq.~(\ref{sm:fegeneral}), which prefers $N_{\rm F}$ control modes $b_{n,m}$ to share the same orientation with the control mode $b_{\rm control}$.

The basic idea is to control different coupling terms with different ancillary modes, and to control the number of active ancillary modes with additional loss channels. For each coupling term $L_{n,m}$ in the original Ising model, we need a pair of frustration-eliminating coupling channels, $L_{\rm ans}^{n,m}$ and $L_{\rm ani}^{n,m}$, which comprises three additional modes, i.e., $b_{n,m}$, $a_{\rm ans}^{n,m}$ and $a_{\rm ani}^{n,m}$. In addition to these ancillary modes distributed to each coupling term, we also need a control mode $b_{\rm control}$ to adjust the number of flipped couplings. Therefore, the relation between the total number of additional modes $N_{\rm ad}$ and the number of coupling terms $N_{\rm c}$ in the original model is
\begin{eqnarray}
N_{\rm ad}=3N_{\rm c}+1.
\end{eqnarray}
\subsection{Semi-classical equations and error identification}
Simulating the quantum model of frustration elimination is in general a challenge. Therefore, we alternatively use the following semi-classical equations under the mean-field approximation,
\begin{eqnarray}
\frac{\partial A_i}{\partial t}\equiv\frac{\partial \langle a_i\rangle}{\partial t}\approx i\langle[H,a_i]\rangle-\sum_k\frac{\gamma_k}{2}\left[\langle[a_i,L^{\dag}_{k}]\rangle\langle L_k\rangle-[a_i,L_{k}]\rangle\langle L_k^{\dag}\rangle\right].
\end{eqnarray}
The equations for all the modes in Eq.~(\ref{sm:fegeneral}) are as follows,
\begin{eqnarray}
\frac{\partial}{\partial t}A_{n}&=&PA_{n}^*-\gamma_{\rm d}A_n|A_n|^2-\gamma_{\rm c}\sum_{n>m}(A_n+C_{n,m}A_m+A_{\rm ans}^{n,m}+A_{\rm ani}^{n,m})\nonumber\\
&&-\gamma_{\rm c}\sum_{n<m}C_{n,m}(A_n+C_{n,m}A_m+A_{\rm ans}^{n,m}+A_{\rm ani}^{n,m}),\nonumber\\
\frac{\partial}{\partial t}B_{n,m}&=&PB_{n,m}^*-\gamma_{\rm d}B_{n,m}|B_{n,m}|^2-\gamma_{\rm c}(B_{\rm control}+B_{n,m}-iA_{\rm ans}^{n,m}+iA_{\rm ani}^{n,m})\nonumber\\
&&-\frac{\gamma_{\rm c}}{2}\left[\sum_{l>m}B_{l,m}+(2N_{\rm F}-N_{\rm c})B_{\rm control}\right],\nonumber\\
\frac{\partial}{\partial t}A_{\rm ans}^{n,m}&=&P{A_{\rm ani}^{n,m}}^*-\gamma_{\rm d}A_{\rm ans}^{n,m}|A_{\rm ani}^{n,m}|^2-\gamma_{\rm c}(A_n+C_{n,m}A_j+iB_{\rm control}+iB_{n,m}+2A_{\rm ans}^{n,m}),\nonumber\\
\frac{\partial}{\partial t}A_{\rm ani}^{n,m}&=&P{A_{\rm ans}^{n,m}}^*-\gamma_{\rm d}A_{\rm ani}^{n,m}|A_{\rm ans}^{n,m}|^2-\gamma_{\rm c}(A_n+C_{n,m}A_j-iB_{\rm n,m}-iB_{n,m}+2A_{\rm ani}^{n,m}),\nonumber\\
\frac{\partial}{\partial t}B_{\rm control}&=&PB_{\rm control}^*-\gamma_{\rm d}B_{\rm control}|B_{\rm control}|^2-\gamma_{\rm c}\sum_{n>m}(B_{\rm control}+B_{n,m}-iA_{\rm ans}^{n,m}+iA_{\rm ani}^{n,m})-\nonumber\\
&&(2N_{\rm F}-N_{\rm c})\gamma_{\rm c}\left[\sum_{l>m}B_{l,m}+(2N_{\rm F}-N_{\rm c})B_{\rm control}\right].
\end{eqnarray}
Here $A_n=\langle a_n\rangle$, $B_{n,m}=\langle b_{n,m}\rangle$, $B_{\rm control}=\langle b_{\rm control}\rangle$, $A_{\rm ani}^{n,m}=\langle a_{\rm ani}^{n,m}\rangle$, and $A_{\rm ans}^{n,m}=\langle a_{\rm ans}^{n,m}\rangle$.

Without frustration, the amplitudes of all modes should have the same absolute value. Therefore, we can use the amplitude inhomogeneity as an indicator for errors. We first define the average amplitude as
\begin{eqnarray}
\bar{A}=\frac{\sum_{n}|A_n|+\sum_{n,m}(|B_{n,m}|+|A_{\rm ans}^{n,m}|+|A_{\rm ani}^{n,m}|)+|B_{\rm control}|}{N+3N_{\rm c}+1}.
\end{eqnarray}
The cumulative fluctuation of the amplitudes can then be defined as
\begin{eqnarray}
F=\sum_{n}(|A_n|-\bar{A})^2+\sum_{n,m}[(|B_{n,m}|-\bar{A})^2+(|A_{\rm ans}^{n,m}|-\bar{A})^2+(|A_{\rm ani}^{n,m}|-\bar{A})^2]+(|B_{\rm control}|-\bar{A})^2.\nonumber\\
\end{eqnarray}
Note that this fluctuation contains the inhomogeneity in both signal modes and ancillary modes.
\subsection{Additional numerical results}
\begin{figure}[h]
\center
\includegraphics[width=6in]{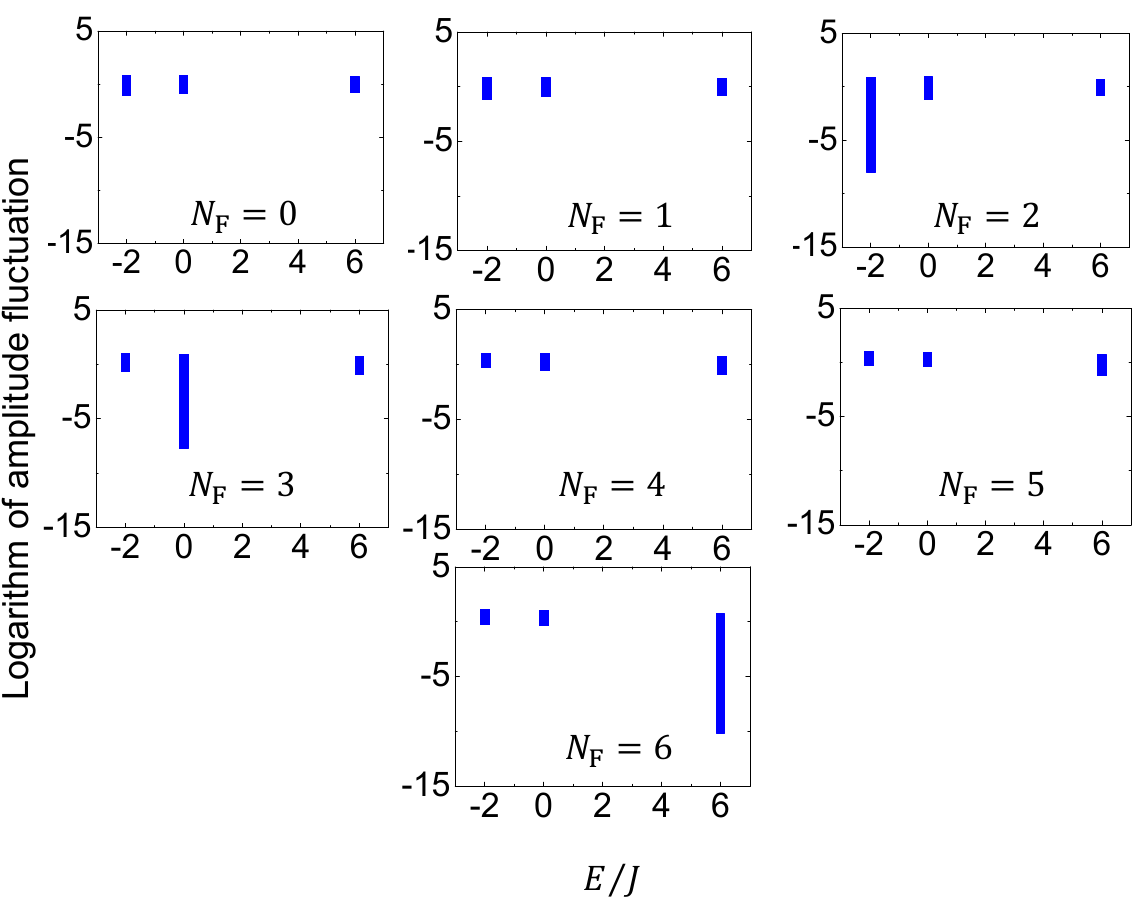}
\caption{Relation between the logarithm of the amplitude fluctuation and the Ising energy of the CIM states obtained with random initial conditions. The Ising model studied is a four-mode one with all-to-all antiferromagnetic coupling. The seven subfigures correspond to seven different numbers of flipped couplings. For $N_{\rm F}=0,1$, no solution is found due to the frustration. The system reaches frustration elimination, and the ground states are found for $N_{\rm F}=2$. By further increasing $N_{\rm F}$, the excited states with two different excited energies are found for $N_{\rm F}=3$ and $N_{\rm F}=6$.}
\label{sm:fig6}
\end{figure}

\begin{figure}[h]
\center
\includegraphics[width=6in]{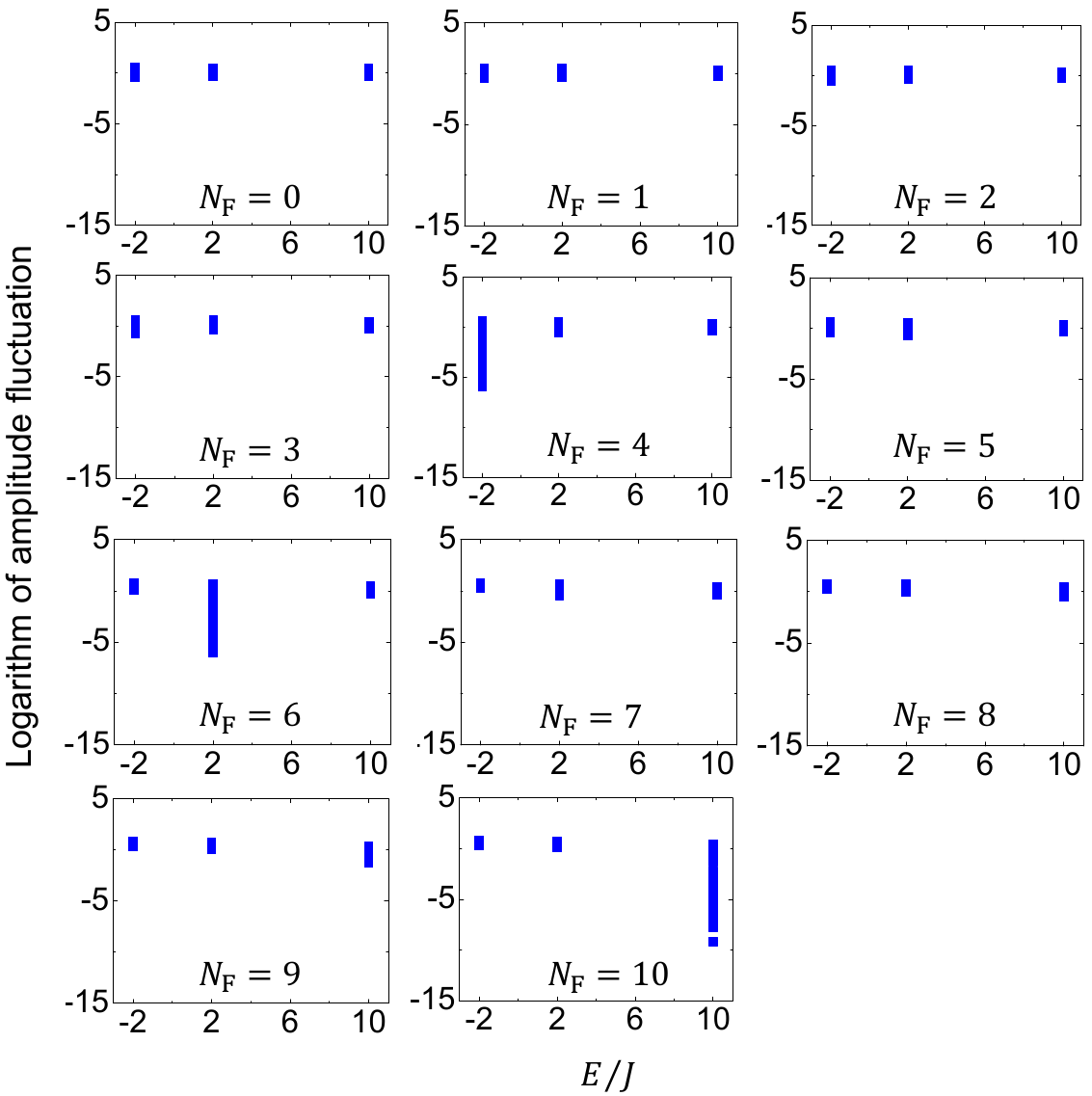}
\caption{Relation between the logarithm of the amplitude fluctuation and the Ising energy of the CIM states obtained with random initial conditions. The Ising model studied is a five-mode one with all-to-all antiferromagnetic coupling. The eleven subfigures correspond to eleven different numbers of flipped couplings. For $N_{\rm F}<4$, no solution is found due to the frustration. The system reaches frustration elimination, and the ground states are found for $N_{\rm F}=4$. By further increasing $N_{\rm F}$, the excited states with two different excited energies are found for $N_{\rm F}=6$ and $N_{\rm F}=10$.}
\label{sm:fig7}
\end{figure}

In the main text, we focus on a three-mode example, which is the simplest example of a frustrated Ising model. Here, we apply the frustration elimination and excited-state search to larger systems, see Figs.~\ref{sm:fig6} and ~\ref{sm:fig7}.

We find that the vanishing fluctuation is also a sufficient condition for correct solutions for the four-mode case and the five-mode case. Therefore, the amplitude inhomogeneity can be used as an indicator for the wrong solutions. In addition, the excited states search is also effective in the examples studied here. By adding more flipped coupling terms, we can find the states with arbitrary energy, which is not possible in current CIMs.

To conclude, we remark that the probability to find correct solutions is reduced due to the additional ancillary modes. However, this reduction can be partially compensated by the self-checking functionality and the potential advantages brought by quantum effects. Presently, the main advantage of these ancillary modes in the semi-classical regime is the search for states with arbitrary energy.
\bibliography{totalreference}